\documentclass{PoS}

\title{Clusters and higher moments of proton number fluctuations}

\ShortTitle{Clusters and higher moments of proton number fluctuations}

\author{\speaker{Boris Tom\'a\v{s}ik}\\
        Univerzita Mateja Bela, Bansk\'a Bystrica, Tajovsk\'eho 40, Slovakia\\
               FNSPE, \v{C}esk\'e vysok\'e u\v{c}en\'i technick\'e v Praze, B\v{r}ehov\'a 7, 
               Prague 1, Czech Republic\\
        E-mail: \email{boris.tomasik@cern.ch}}

\author{Zuzana Paul\'inyov\'a\\
        Univerzita Pavla Jozefa \v{S}af\'arika, Jesenn\'a 5, Ko\v{s}ice, Slovakia\\
        }

\author{Jan Steinheimer\\
        Frankfurt Institute for Advanced Studies, Johann-Wolfgang-Goethe-Universit\"at, 
        Ruth-Moufgang-Stra{\ss}e 1, Frankfurt, Germany\\
        }

\author{Marcus Bleicher\\
        Frankfurt Institute for Advanced Studies, Johann-Wolfgang-Goethe-Universit\"at, 
        Ruth-Moufgang-Stra{\ss}e 1,Frankfurt, Germany\\
        Institut f\"ur theoretische Physik, Goethe-Universit\"at Frankfurt, 
        Max-von-Laue-Stra{\ss}e 1, Frankfurt, Germany\\
        GSI Helmholtzzentrum für Schwerionenforschung GmbH, Planckstra{\ss}e 1,  Darmstadt, Germany
        }

\abstract{Production of deuterons by coalescence  has an important influence on the moments
of the observed proton number distribution. Therefore, this effect must be taken into account 
when physics conclusions about baryon number fluctuation are drawn from the measurement 
of proton number fluctuations. We also show that a 
measurement of the third and fourth moments of the
deuteron number distribution would allow to distinguish whether deuterons are produced statistically 
or by coalescence. }

\FullConference{Critical Point and Onset of Deconfinement - CPOD2017\\
		7-11 August, 2017\\
		The Wang Center, Stony Brook University, Stony Brook, NY}

\begin{document}

\section{Introduction}

Moments of the proton number distribution are measured in ultrarelativistic nuclear collisions up to 
fourth order. The prime reason is the study of the QCD phase diagram, because they are used as a proxy 
for the higher order baryon number susceptibilities. The purpose of this paper is twofold. Firstly, we 
 investigate how the skewness and kurtosis of the proton number distribution are influenced by coalescence 
 which takes some  of the protons and puts them into deuterons. Secondly, we point out that the 
 measurement of the higher moments of deuteron number distribution can help to resolve the mechanism 
 of deuteron production, whether it is coalescence or direct statistical production. 

Higher order susceptibilities with respect to baryon number are being calculated on the lattice 
with increasing accuracy, as we have also witnessed at this conference \cite{Bazavov, Gunther}.
They can be used for a more precise mapping of the QCD phase diagram because their values
are very sensitive to the vicinity of the crossover from the confined to the deconfined state. For practical 
measurement they are related to the higher moments of the  (net) baryon 
number distribution. Unfortunately, baryon number is actually not measurable at least because one
cannot detect and count neutrons (and antineutrons). Therefore, the measurement must be based on 
protons only. Arguments exist that the proton number measurement is a good proxy for the 
full measurements of the baryon number \cite{kitazawa1,kitazawa2}. However, the measured
results for the moments of the proton number distribution may be influenced by other effects. 
Among them there are the conservation of baryon number, final state hadronic interactions
\cite{Steinheimer}, as well as issues like detector efficiency and acceptance limitations. It is important 
to realise that by going to as high as third and fourth moments of a number distribution the 
measurement becomes sensitive to rather fine details of the distribution. 

Production of deuterons is among the processes which can affect the observed number of protons.
With the binding energy of 2.2~MeV a deuteron is too fragile to survive in an environment with 
the temperature of more than 100~MeV. Thus it is widely believed that the deuterons are made by 
coalescence of protons and neutrons. Their average number is usually well reproduced if one 
assumes that it is given as $Bn_p^2$, where $n_p$ is the number of protons and $B$ is the coalescence
factor. Here we will push the idea of proportionality further and assume that in \emph{each} event the number 
of deuterons is proportional (up to fluctuations) to the actual number of protons to the second power. 
Subtracting in each event the protons which disappear in deuterons from the number of all protons 
will modify the number distribution of observed protons. At low energy deuteron production is 
a non-negligible effect. However, even if this effect becomes small at higher energies, it still could 
show up in the higher moments of proton distribution. 
We have formulated a model in this spirit and calculated the expected influence on the skewness and 
kurtosis of the proton number distribution. 

Coming back to the mechanism of deuteron production, higher moments of their number distribution 
could help resolve the question how the deuterons are actually produced. Although the coalescence 
scenario seems quite natural, the mean yields can also be interpreted in terms of the statistical 
model. Clear difference between the two mechanisms of deuteron production appears in higher 
moments of deuteron number distribution. Statistical model should lead to Poissonian number 
distribution while coalescence couples the deuteron number distribution with that of protons (and 
neutrons).  Once we have formulated the model of how this coupling is actually realised it is 
straightforward to use it not only for proton number fluctuations but also for the description of 
deuteron number fluctuations. We shall see the differences between results of the two models.


\section{The observables}

In this paper we will evaluate moments of the proton number distributions. In the second part 
we shall also study the distribution of the deuteron number $n_d$. 
Averages over events are denoted with angular brackets, so that $\langle n_p \rangle$ 
and $\langle n_d \rangle$ stand for the mean proton number and mean deuteron  number, 
respectively. 

We shall look at the variance
\begin{equation}
\sigma^2 = {\left\langle \left ( n_p - \langle n_p \rangle \right  )^2\right \rangle}
\label{e:sigma}
\end{equation}
the skewness
\begin{equation}
S = \frac{{\left\langle \left ( n_p - \langle n_p \rangle \right  )^2\right \rangle}}{\sigma^3}
\end{equation}
and the kurtosis
\begin{equation}
\kappa = \frac{\left\langle \left ( n_p - \langle n_p \rangle \right  )^2\right \rangle}{\sigma^4} - 3
\label{e:kappa}
\end{equation}
of the number distributions.


\section{The model}
\label{s:formulas}

We formulate here the simplest model which takes into account that deuteron number scales 
with the square of the proton number. 

We start by assuming that the number of all initial protons $n_i$ fluctuates according to a Poisson distribution
with the mean $\lambda_p$
\begin{equation}
\label{e:mip}
P_i(n_i) = \lambda_p^{n_i} \frac{e^{-\lambda_p}}{n_i!}\,  .
\end{equation}
It is important to realise that $n_i$ is not measurable, since it also includes those protons which become 
parts of deuterons. Thus also $\lambda_p$ is not directly observable and will have to be obtained from common 
analysis of proton and deuteron yields. 

Production of deuterons is random process, as well.  We assume that the actual number of deuterons 
in every event is drawn from a Poisson distribution. The mean of that distribution in an event with $n_i$
initial protons is 
\begin{equation}
\label{e:lp2}
\lambda_d = Bn_i^2\, ,
\end{equation}
as dictated by the coalescence model. The coalescence parameter $B$ is to be determined from 
experimental data.
Then, the \emph{conditional} probability to have $n_d$ deuterons in an event with $n_i$ initial protons
is 
\begin{equation}
P_d(n_d|n_i) = \lambda_d^{n_d} \frac{e^{-\lambda_d}}{n_d!} = 
\left ( Bn_i^2 \right )^{n_d} \frac{e^{-Bn_i^2}}{n_d!}\,  .
\label{e:dcond}
\end{equation}

Combining this distribution with that of initial protons given by eq.~(\ref{e:mip}) we can obtain the 
distribution for the number of \emph{observed} protons $n_p$. We recall that those protons are 
observed which do not disappear from the balance as parts of deuterons. Thus
\begin{equation}
n_p = n_i - n_d\,  ,
\end{equation}
and the corresponding distribution is obtained as
\begin{equation}
P(n_p) = \sum_{n_i \ge n_p} P_i(n_i) P_d(n_i - n_p | n_i)\,  .
\end{equation} 
The sum on the right-hand side can be evaluated and thus we know the whole distribution. From 
this all moments can be calculated via equations (\ref{e:sigma}-\ref{e:kappa}). 

In order to obtain the observed number of deuterons we have to sum the conditional probability
$P_d(n_d|n_i)$ for all possible values of $n_i$. Thus we arrive at the folding
\begin{equation}
\label{e:ddsimple}
P_d(n_d) = \sum_{n_i \ge n_d} P_d (n_d | n_i) P_i(n_i)\,   .
\end{equation}
This distribution can be evaluated straightforwardly. Again, all its moments can be calculated, 
as well.

\begin{figure}[t]
\begin{center}
\includegraphics[width=0.7\textwidth]{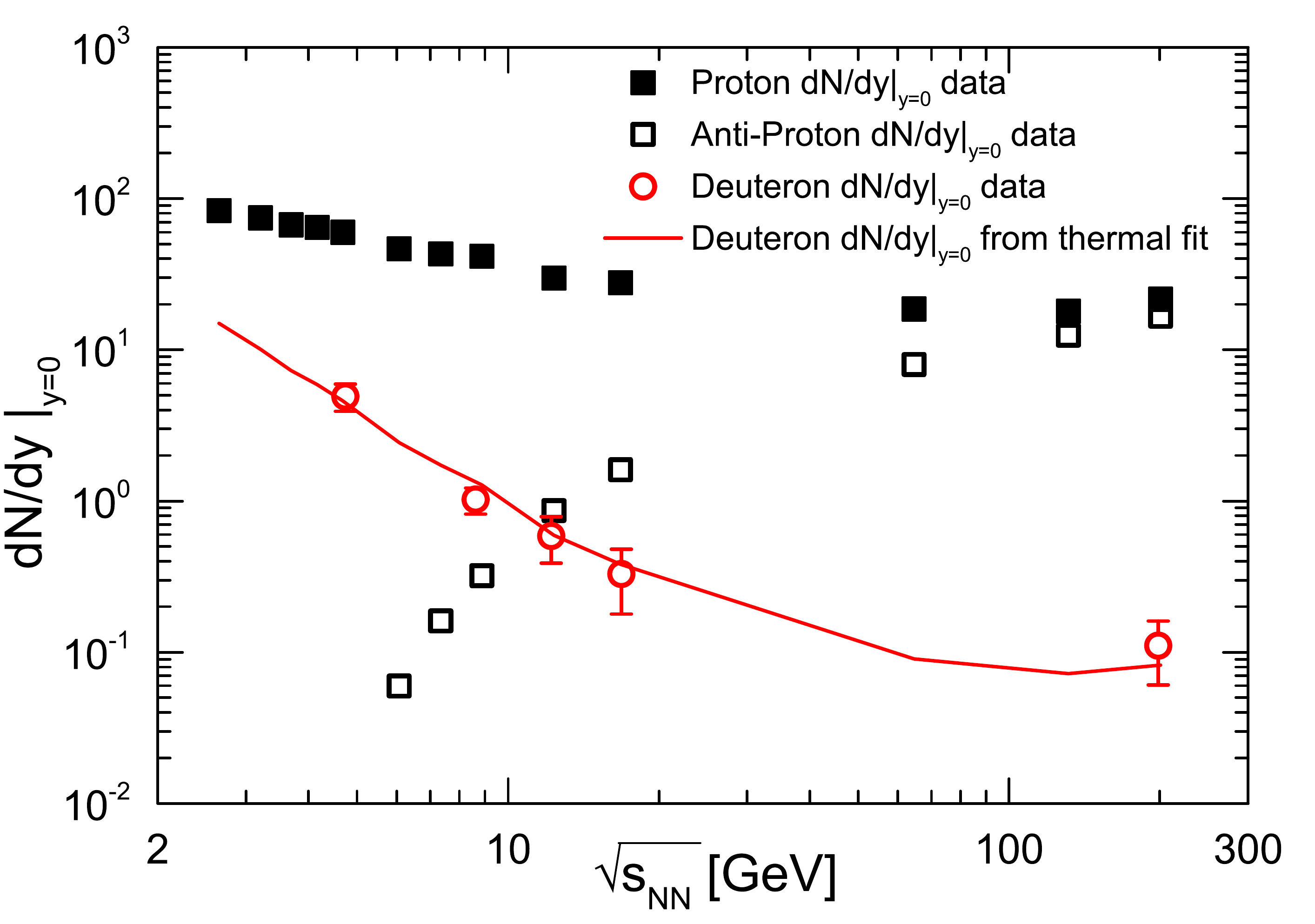}
\end{center}
\caption{
Fit to the deuteron yields at midrapidity based on the parametrisation
(\ref{e:d2p}). Experimental data from Au+Au collisions are plotted for protons, antiprotons
\cite{PBM1}, and deuterons \cite{PBM2}.
}
\label{f:determB}
\end{figure}

In summary, we have two parameters in our model: the mean initial number of protons $\lambda_p$ 
and the coalescence parameter $B$. They have to be determined separately for every collision energy.
The two observables that are used to determine them are the mean observed proton number and the mean 
deuteron number
\begin{eqnarray}
\langle n_p \rangle & = & \sum_{n_p} n_p P_p(n_p) \\
\langle n_d \rangle & = & \sum_{n_d} n_d P_d(n_d)\,  .
\end{eqnarray}
In order to get the deuteron number at all energies that we need, we have fitted the collision energy 
dependence of the deuteron-to-proton ratio 
\begin{equation}
\frac{\langle n_d \rangle}{\langle n_p \rangle} = 
0.8 \left [ \frac{\sqrt{s_{NN}}}{1\,\mathrm{GeV}} \right ]^{-1.55} + 0.0036\,   .
\label{e:d2p}
\end{equation}
The quality of the fit can be inspected from Fig.~\ref{f:determB}.

Using these formulas we have calculated the higher moments of the proton number distribution. 
At higher collision energies, antiprotons also have to be taken into account. In principle, they can also make
the antideuterons and this is expressed via the same type of distributions as we have had for protons 
and deuterons. In this case, however, we made Monte Carlo simulations of proton and antiproton 
production according to the formulated model. At lower energies we have checked that the MC simulation 
and the direct calculation lead to the same results.


\section{Results for net proton number fluctuations}

\begin{figure}
\begin{center}
\includegraphics[width=0.48\textwidth]{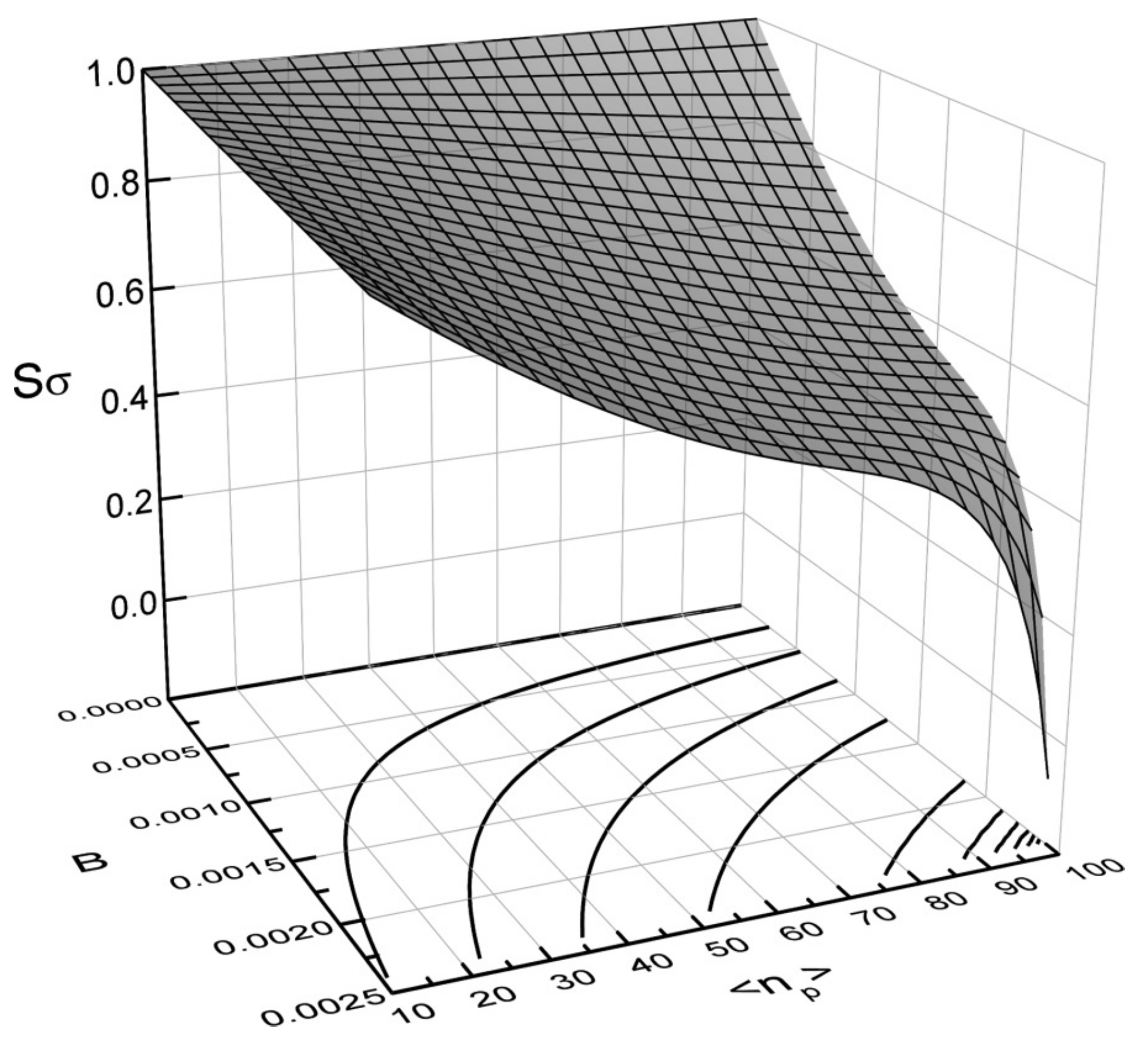}
\includegraphics[width=0.48\textwidth]{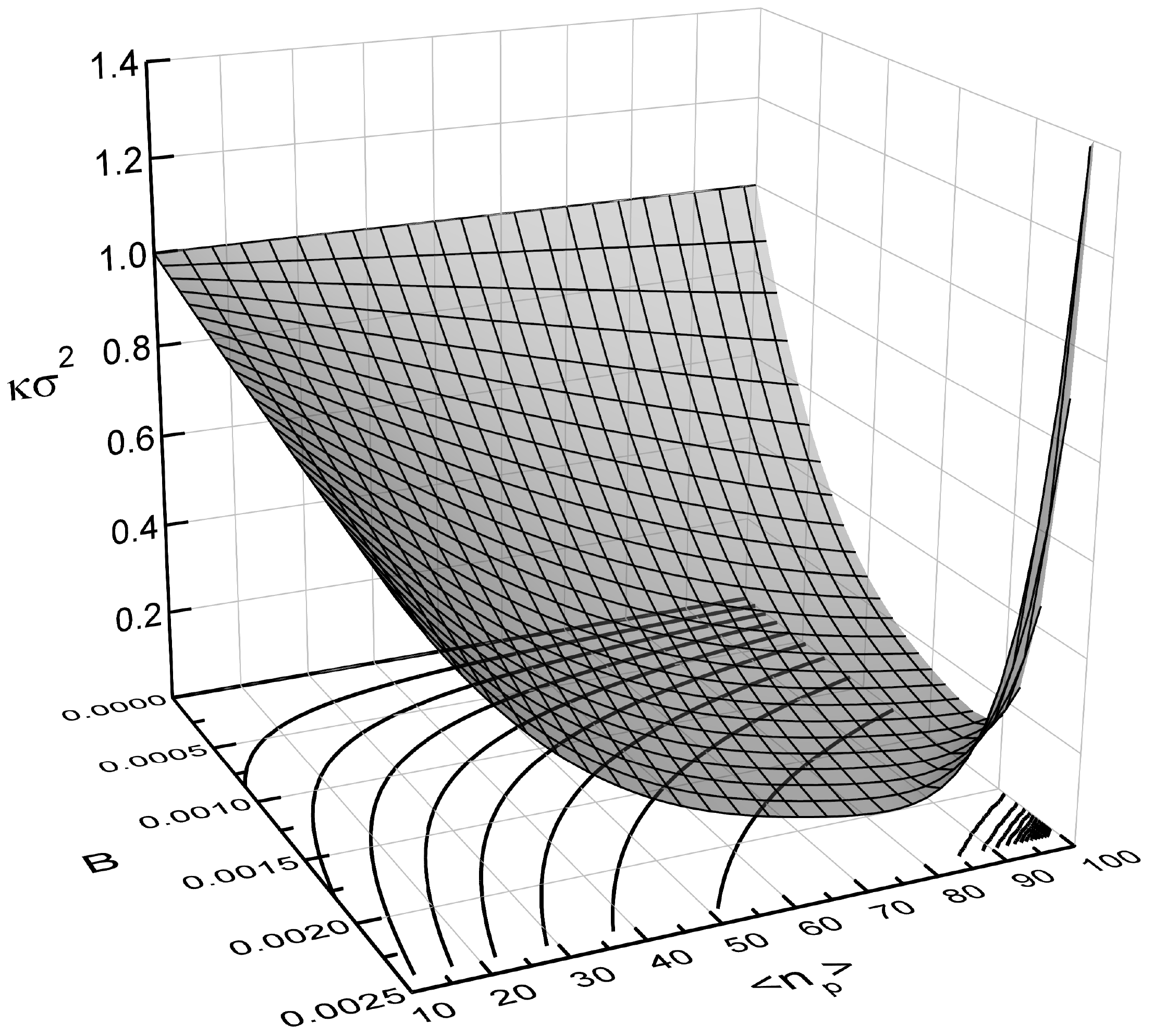}
\end{center}
\caption{Scaled skewness $S\sigma$ (left) and scaled kurtosis $\kappa\sigma^2$
(right) as functions of mean
observed proton number $\langle n_p \rangle$ and the coalescence parameter $B$. 
}
\label{f:SkewKurt}
\end{figure}

With the help of the derived model we have determined the values of the scaled skewness 
$S\sigma$ and scaled kurtosis $\kappa\sigma^2$ as functions of the mean observed proton number
$\langle n_p \rangle $ and the coalescence parameter $B$. The results are plotted in 
Fig.~\ref{f:SkewKurt}.
%
%
First of all we see that if no deuterons are produced, i.e.\ at $B = 0$, we recover the trivial values for the Poisson distribution: $S\sigma = \kappa\sigma^2 = 1$. On the other hand, many deuterons are produced if 
the number of protons is large or for large value of the coalescence parameter $B$. Then the scaled moments
are modified dramatically. In most extreme cases they go down as low as to 0.2. The values of $\langle n_p\rangle$ 
and $B$ in that figure are realistic and the highest values can be attained in nuclear collisions at lower 
BNL-AGS energies. We thus stress that the effect of the deuteron production on the shape of the 
proton number distribution may play an important role and definitely must be taken into account. 

\begin{figure}[t]
\begin{center}
\includegraphics[width=0.49\textwidth]{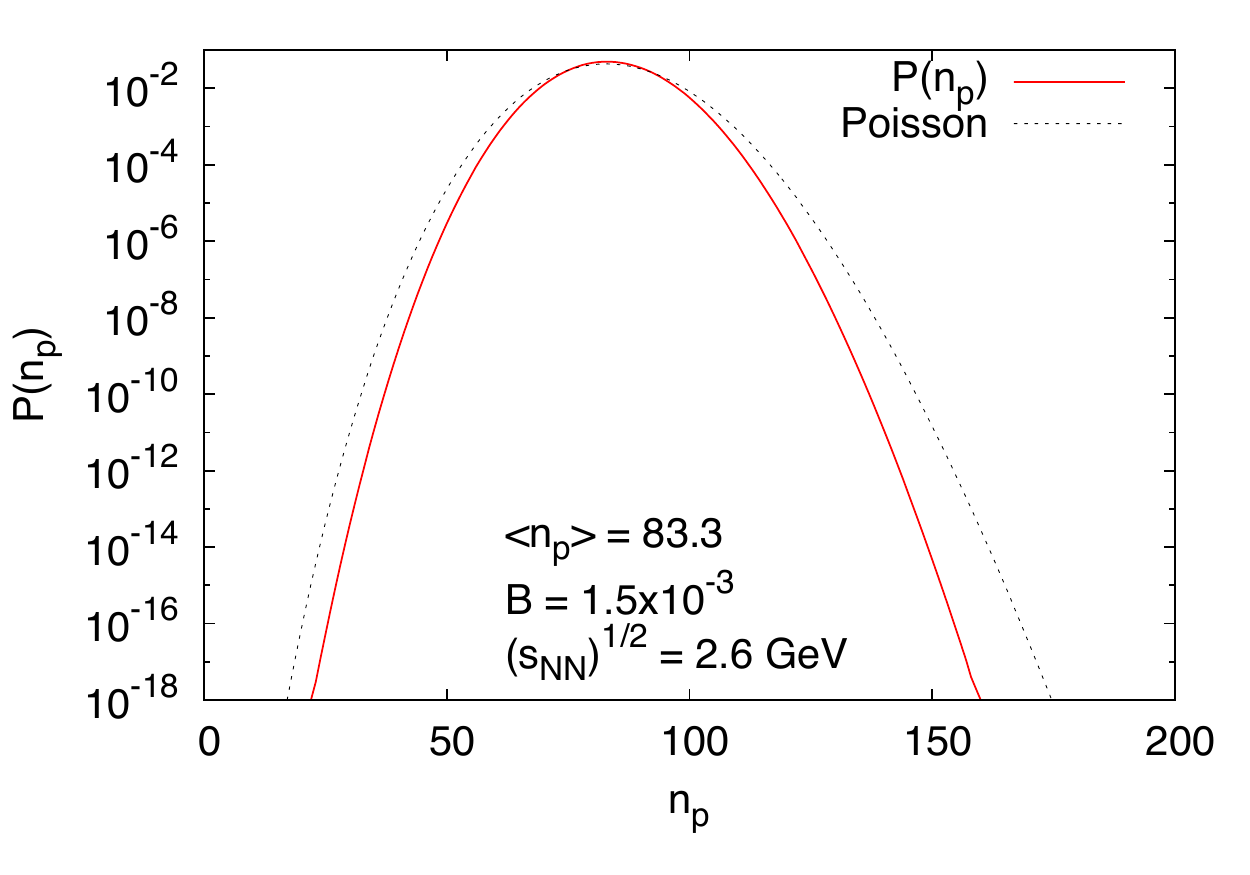}
\includegraphics[width=0.49\textwidth]{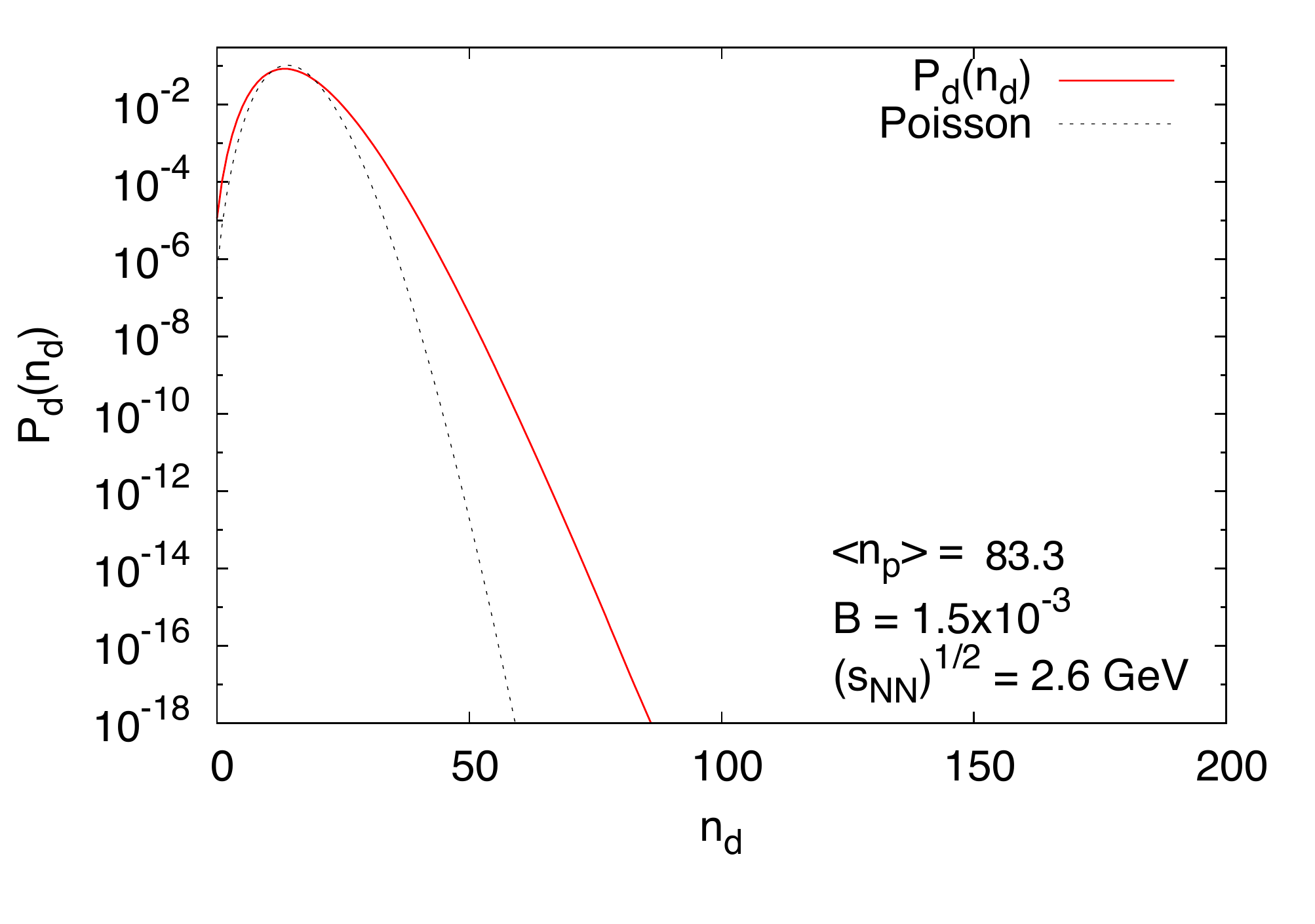}
\end{center}
\caption{
Number distribution of observed protons (left) and of deuterons (right) calculated for 
$\langle n_p\rangle = 83.4$
and $B=1.6\times 10^{-3}$. These parameter values describe proton and deuteron production 
from Au+Au collisions at $\sqrt{s_{NN}} = 2.6$~GeV. For comparison, dashed curves show the Poisson 
distribution with the same mean value as the calculated distributions.  
}
\label{f:numdist}
\end{figure}

In order to better understand where the large modification of the values comes from, we have also plotted 
the proton and deuteron number distributions (Fig.~\ref{f:numdist}). 
%
%
They have been calculated for values of $\langle n_p\rangle$ and $B$ relevant for central Au+Au collisions
at $\sqrt{s_{NN}} = 2.6$~GeV. In general, we observe that for the proton number distribution, production 
of deuterons cuts from the high-multiplicity end because there coalescence is more pronounced, since 
it is proportional to the square of the proton number. Hence, in general, the observed proton number 
distribution is narrower than the Poisson. The overall effect is that the resulting $P_p(n_p)$ has smaller 
tails and is more skewed towards lower proton numbers than the Poisson distribution. This is quantified
by the calculated skewness and kurtosis. 

On the obtained deuteron distribution the effect is just opposite. Since the mean deuteron number is 
proportional to $n_i^2$, the high multiplicity tail of the distribution is enhanced and really heavy. 
It is clearly heavier than the reference Poisson distribution. This will result in significantly larger values for the 
skewness and kurtosis later in this paper. 

\begin{figure}
\begin{center}
\includegraphics[width=0.48\textwidth]{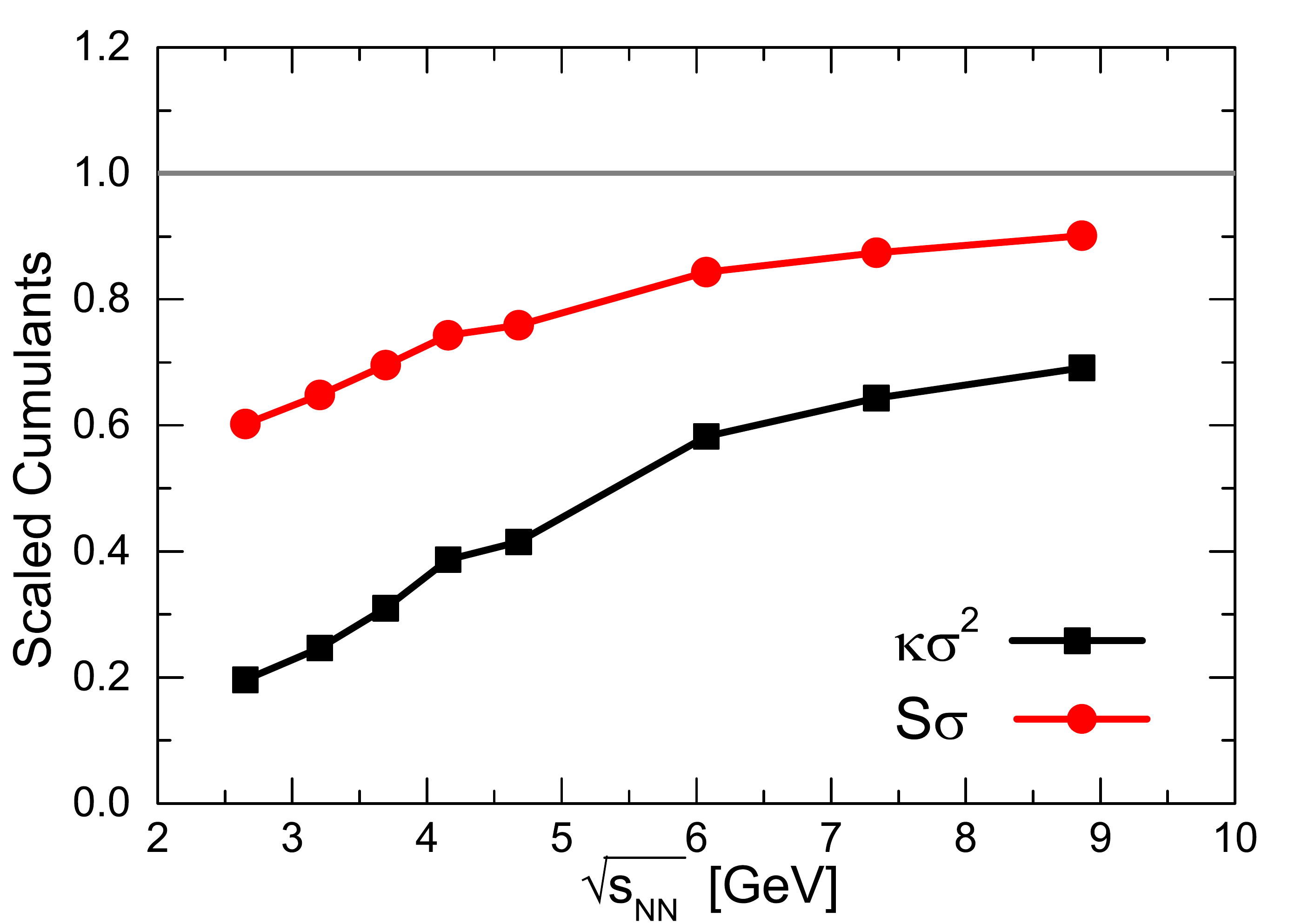}
\includegraphics[width=0.48\textwidth]{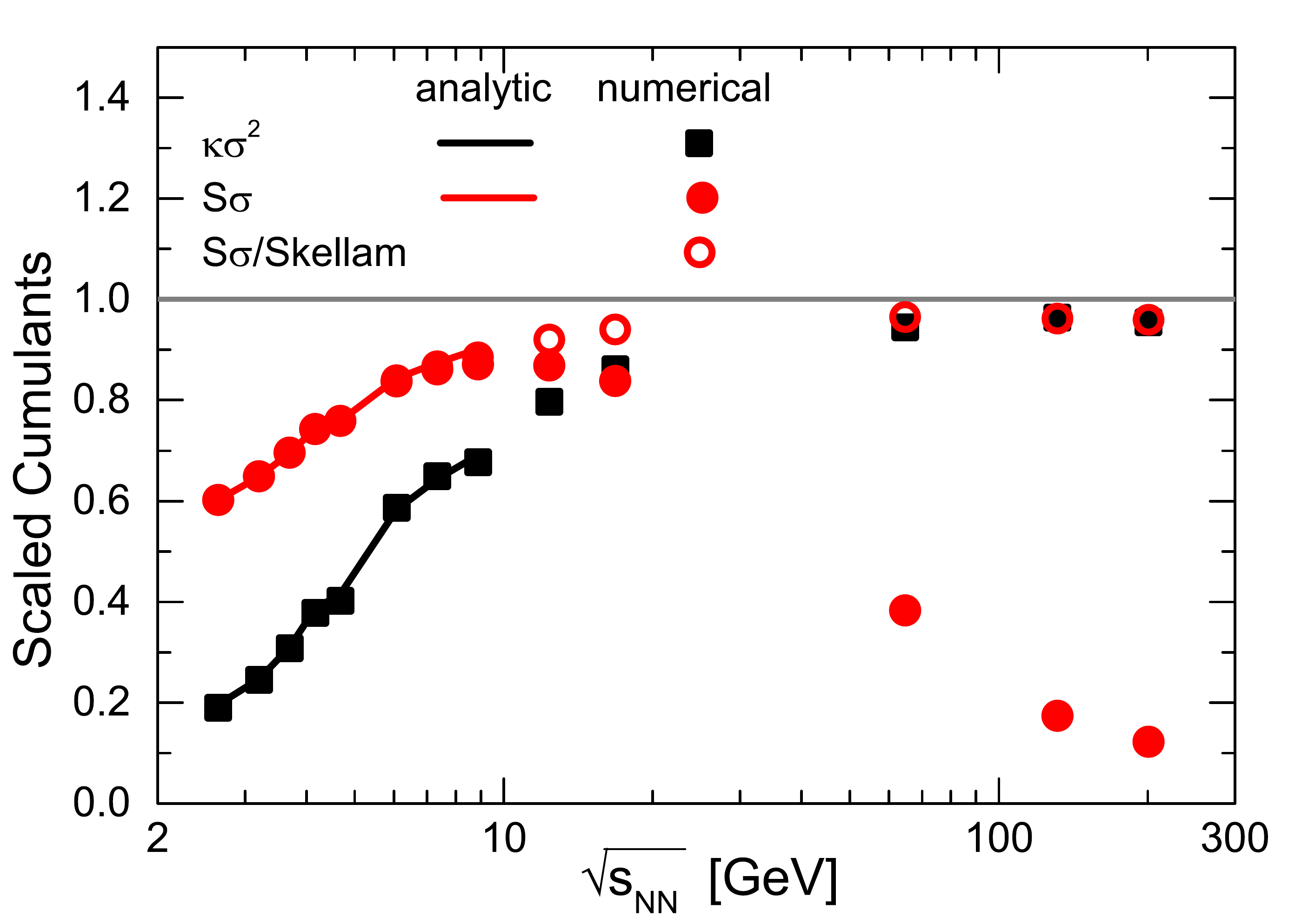}
\end{center}
\caption{Scaled skewness $S\sigma$ and scaled kurtosis $\kappa\sigma^2$ as functions
of collision energy for central Au+Au collisions. In the left panel no antiprotons are taken 
into account and results are calculated with the help of formulas from Section~\ref{s:formulas}. 
In the right panel antiproton production has  been taken into account and results are obtained 
from Monte Carlo simulations. 
}
\label{f:resEdep}
\end{figure}

After we have understood the results qualitatively, we are ready to plug in phenomenologically relevant 
values of the parameters and calculate the collision energy dependence of the skewness and kurtosis. Recall 
that we want to study the fluctuations of the net proton number. However, at energies below 
$\sqrt{s_{NN}}\approx 10$~GeV antiproton production can be neglected and we can just look at 
proton number distribution. In this interval we have calculated the higher moments with the formulas
from Section~\ref{s:formulas}. As a cross-check, $S\sigma$ and $\kappa\sigma^2$ have been 
determined by Monte Carlo simulations for all collision energies. The results are plotted in Fig.~\ref{f:resEdep}.
%

When also antiprotons are produced, Skellam distribution is expected if both proton and antiproton
numbers follow the Poisson distribution. Therefore, for the skewness we also plot the ratio of our 
result to the expectation due to Skellam distribution. This is important. Our results show a
dramatic decrease of $S\sigma$ as the collision energy increases to top RHIC energies of 
130 and 200 GeV pre nucleon pair. However, when divided by the value due to Skellam distribution
we see that the ratio converges towards 1, so that the result is trivial.  
Nevertheless, we see that the results for collision energies below 10~GeV per nucleon pair are
strongly influenced by formation of deuterons.


\section{Thermal production vs coalescence of deuterons}

If deuterons are produced simply according to the statistical model, then their number distribution 
should be Poissonian and $S\sigma = \kappa\sigma^2 = 1$. As it was illustrated in 
Fig.~\ref{f:numdist} right, coalescence leads to skewness and kurtosis which are much increased 
with respect to that expectation. However, we note that this result was based on rather strong
assumption that the mean number of deuterons scales with $n_i^2$. As a matter of fact, 
deuterons are bound states of a proton and a neutron and so it would be more natural to 
expect that 
\begin{equation}
\lambda_d = B n_i n_j\,   ,
\label{e:dpn}
\end{equation}
where $n_j$ is the initial number of all neutrons, analogically to initial number of all protons $n_i$. 
The number of neutrons fluctuates according to Poisson distribution for which we assume the same 
mean value as we did for initial proton number distribution. The conditional probability that $n_d$ 
deutrons are observed in case where $n_i$ initial protons and $n_j$ initial neutrons were produced
is
\begin{equation}
P(n_d|n_i,n_j) = \lambda_d^{n_d} \frac{e^{-\lambda_d}}{n_d!} = 
\left ( Bn_in_j \right )^{n_d} \frac{e^{-Bn_in_j}}{n_d!}\,  .
\label{e:dcondpn}
\end{equation}
Consequently, the distribution of deuteron number is 
\begin{equation}
\label{e:dfullpn}
P_d(n_d) = \sum_{n_i,n_j\ge n_d} P_d(n_d|n_i,n_j) P_i(n_i)P_j(n_j)\,  .
\end{equation}
For later reference, we call the two models that we have formulated here as 
\begin{description}
\item[Model A] given by eqs.~(\ref{e:lp2}),  (\ref{e:dcond}), and (\ref{e:ddsimple});
\item[Model B] given by eqs.~(\ref{e:dpn}), (\ref{e:dcondpn}), and (\ref{e:dfullpn}).
\end{description}


\section{Results for deuteron number fluctuations}

\begin{figure}[t]
\begin{center}
\includegraphics[width=0.75\textwidth]{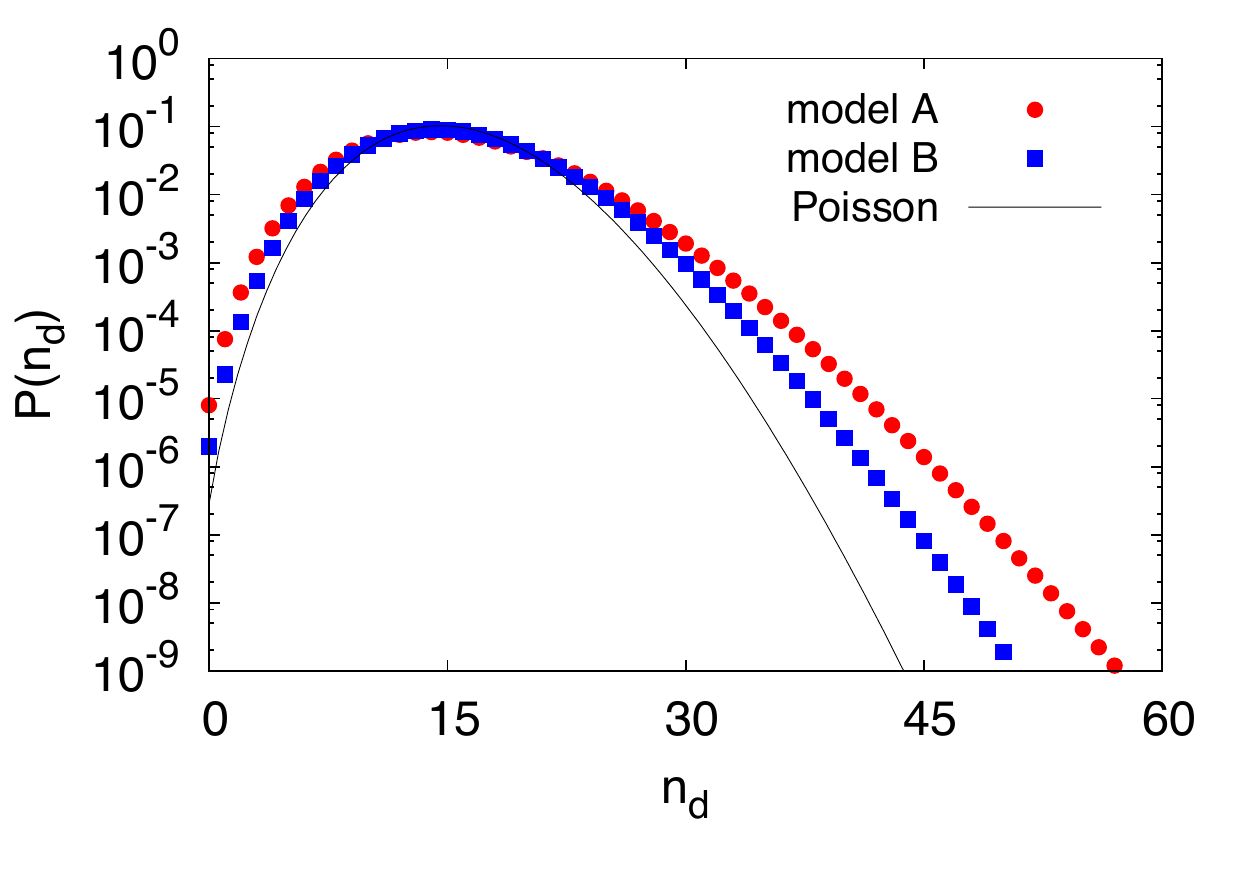}
\end{center}
\caption{
Deuteron number distributions determined for model parameters which correspond to 
Au+Au collisions at $\sqrt{s_{NN}} = 2.6$~GeV. All distributions have the same mean. 
}
\label{f:d-distr}
\end{figure}

We first inspect the difference between the individual models in Fig.~\ref{f:d-distr}.
%
It shows the deuteron number distributions with parameters suitable for the description of 
proton and deuteron production in Au+Au collisions at $\sqrt{s_{NN}} = 2.6$~GeV.
It is seen that both Model A and Model B lead to number distributions which are clearly 
more skewed and have heavier tails than the reference with the same mean which is provided by 
the Poisson distribution. While Poisson distribution leads to 
$\sigma^2/\langle n_d\rangle = S\sigma = \kappa\sigma^2 = 1$,
we obtain for Model A: $\sigma^2/\langle n_d\rangle = 1.609$, 
$S\sigma = 2.218$, $\kappa\sigma^2 = 6.915$;
and for Model B: 
$\sigma^2/\langle n_d\rangle = 1.308$, $S\sigma = 1.616$, $\kappa\sigma^2 2 = 3.422$.

\begin{figure}[t]
\begin{center}
\includegraphics[width=0.48\textwidth]{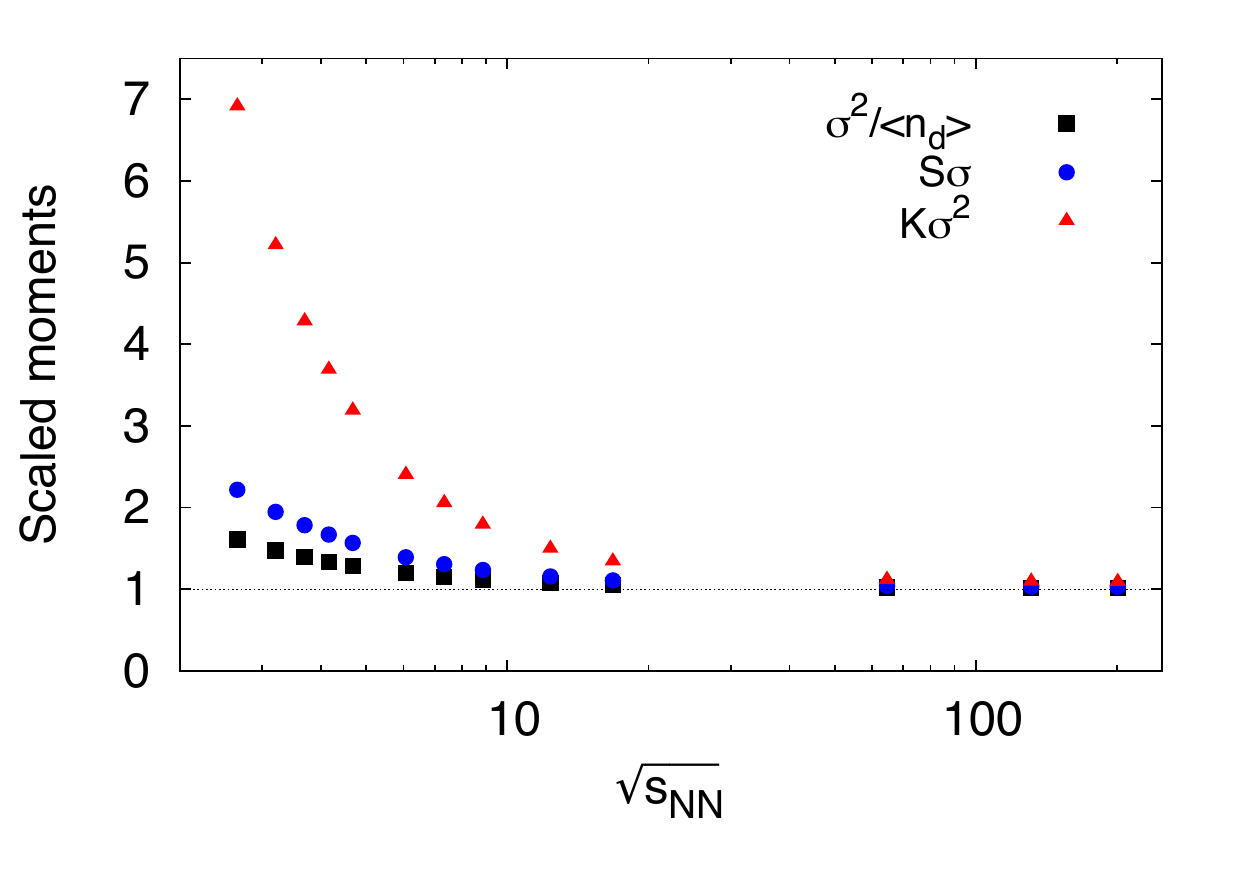}
\includegraphics[width=0.48\textwidth]{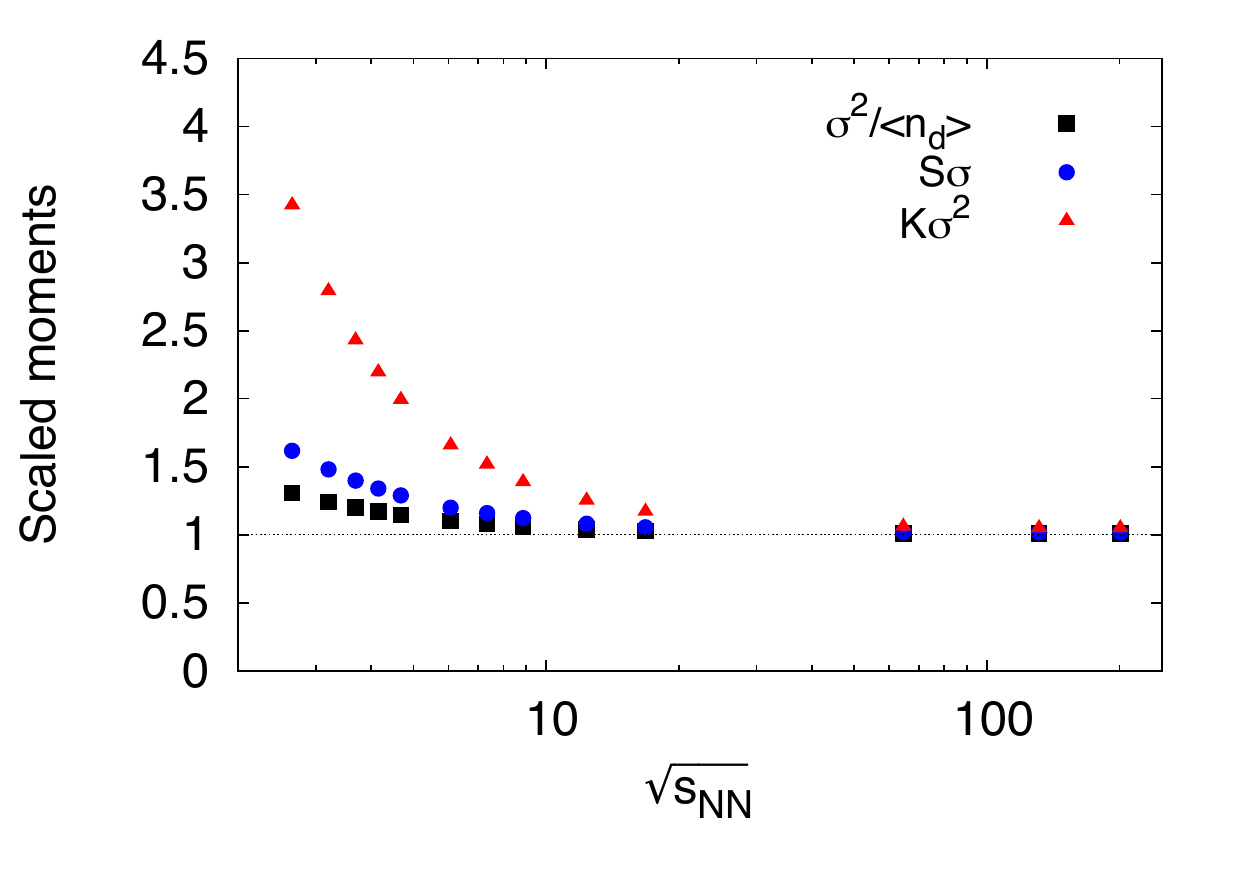}
\end{center}
\caption{
Collision energy dependence for $\sigma^2/\langle n_d\rangle$, $S\sigma$, and $\kappa\sigma^2$
for the model with deuteron production depending on proton number squared (Model A, left) and 
the model assuming independent proton and neutron number fluctuations (Model b, right).
}
\label{f:s-moments}
\end{figure}

Encouraged by this result we can now determine $P_d(n_d)$ for all collision energies and 
calculate the higher  moments. 
The results are plotted in Fig.~\ref{f:s-moments} for Model A (left) and Model B (right).
%
%
In general, we can summarise that going to lower energies makes the effect of coalescence more
visible. It is also more visible in the higher moments of the distribution. The difference with respect 
to Poissonian result is visible in case of both Model A a Model B, although it is about twice as big 
for Model A than it is for Model B. 
We thus offer
a good motivation for  a measurement of higher moments of deuteron distribution.


\section{Conclusions}

Proton number fluctuations are nowadays being widely studied because of their connection to baryon number
fluctuations. 

Here we have shown that the production of deuterons must be taken into account if any physics 
conclusions are to be drawn from the measurement. The disappearance of some protons 
into deuterons has an important effect on the proton number distribution \cite{feckova1}.

We have also shown that by studying higher moments of the deuteron number distribution 
one will be able to decide if the deuterons are produced according to the statistical model or 
if their production is rather described by coalescence \cite{feckova2}. 


\paragraph{Acknowledgments}
We acknowledge the support by VEGA 1/0469/15 (Slovakia).
The work has been supported by the grant 17-04505S of the Czech Science Foundation (GA\v{C}R).
The collaboration has been facilitated by the DAAD PPP project and by the COST Action CA15213 THOR.


\end{document}